\documentclass[prl,twocolumn,superscriptaddress,showpacs]{revtex4}
\usepackage{graphicx,amsmath,amssymb,bm}

\def\be{\begin{equation}}
\def\ee{\end{equation}}
\def\ba{\begin{eqnarray}}
\def\ea{\end{eqnarray}}
\def\bas{\begin{eqnarray*}}
\def\eas{\end{eqnarray*}}

\begin{document}

\title{Medium-mass nuclei from chiral nucleon-nucleon interactions}

\author{G.~Hagen} 
\affiliation{Physics Division, Oak Ridge National
Laboratory, Oak Ridge, TN 37831, USA} 
\author{T.~Papenbrock}
\affiliation{Department of Physics and Astronomy, University of
Tennessee, Knoxville, TN 37996, USA} 
\affiliation{Physics Division,
Oak Ridge National Laboratory, Oak Ridge, TN 37831, USA}
\author{D.J.~Dean} 
\affiliation{Physics Division, Oak Ridge National
Laboratory, Oak Ridge, TN 37831, USA} 
\author{M.~Hjorth-Jensen}
\affiliation{Department of Physics and Center of Mathematics for
Applications, University of Oslo, N-0316 Oslo, Norway}

\begin{abstract}
We compute the binding energies, radii, and densities for selected
medium-mass nuclei within coupled-cluster theory and employ the
``bare'' chiral nucleon-nucleon interaction at order N$^3$LO. We find
rather well-converged results in model spaces consisting of 15
oscillator shells, and the doubly magic nuclei $^{40}$Ca, $^{48}$Ca,
and the exotic $^{48}$Ni are underbound by about 1~MeV per nucleon
within the CCSD approximation. The binding-energy difference between
the mirror nuclei $^{48}$Ca and $^{48}$Ni is close to theoretical mass
table evaluations.  Our computation of the one-body density matrices
and the corresponding natural orbitals and occupation numbers provides
a first step to a microscopic foundation of the nuclear shell model.

\end{abstract}


\maketitle

{\it Introduction.} Ab-initio nuclear structure calculations have made
great progress in the past decade.  Light nuclei up to carbon or so
can now be described in terms of their nucleonic degrees of freedom
and realistic nucleon-nucleon (NN) forces (i.e. those that include
pion exchange and fit the NN phase shifts up to 350~MeV lab energy
with a $\chi^2\approx 1$ per datum) augmented by a three-nucleon force
(3NF)~\cite{Piep01,Nav00,Kam01}. One of the major advances is due to
the systematic construction of nuclear forces within chiral effective
field theory (EFT)~\cite{Weinberg,Kolck94}. In this EFT, unknown
short-ranged physics of the nuclear force is systematically
parametrized in terms of contact terms and their low-energy constants,
while the long-range part of the interaction stems from pion
exchange. One of the hallmarks of this approach is the ``power
counting'', i. e. an expansion of the nuclear Lagrangian in terms of
the momentum ratio $Q/\Lambda$. Here, $Q$ denotes the typical momentum
scale at which the nucleus is probed, while $\Lambda$ denotes the
high-momentum cutoff scale that limits the applicability of the
EFT. Within this approach, three-nucleon forces appear naturally at
order $(Q/\Lambda)^3$, and four-nucleon forces appear at order
$(Q/\Lambda)^4$~\cite{Epel00,N3LO,N3LOEGM}.

The chiral interactions have been probed in light systems up to mass
13~\cite{Epel02,Caurier04,Nogga06,Nav07}.  Fujii~{\it et al.}  have
employed chiral NN interactions for studies of $^{16}$O~\cite{Fujii04}
within the unitary-model-operator approach (UOMA). Unfortunately,
virtually nothing is known about chiral interactions in heavier
nuclei. In particular, a study of their saturation properties is
missing, and the contributions of chiral NN interactions to nuclear
binding and structure in medium-mass nuclei needs to be determined. It
is the purpose of the present Letter to fill this gap.

{\it Ab initio} methods began to explore medium-mass nuclei only very
recently. Gandolfi {\it et al.}~\cite{Gandolfi} employed the auxiliary
field diffusion Monte Carlo method for a computation of the binding
energy of $^{40}$Ca. However, this impressive calculation is not
entirely realistic since the employed Argonne $v'_6$ potential lacks
the spin-orbit interaction. Roth and Navr{\'a}til~\cite{Roth07} employed
softer renormalized NN interactions and computed the binding energy of
$^{40}$Ca within an importance truncated no-core shell model
approach. However, this calculation was
criticized~\cite{comment,CCbench} for its convergence properties, the
violation of Goldstone's linked cluster theorem and the corresponding
lack of size extensivity. In this Letter, we employ the ``bare''
chiral NN interaction~\cite{N3LO} and employ the size extensive
coupled-cluster method~\cite{Coe58,Coe60,Ciz66,Ciz69,Kuem78,Bar07} for
the computation of various properties of the medium-mass nuclei
$^{40}$Ca, $^{48}$Ca, and the exotic $^{48}$Ni. The use of the ``bare''
NN interaction has the advantage that it avoids the introduction of
additional many-body forces that are typically generated in secondary
renormalization procedures of the two-body force.  While our
calculation includes the chiral NN interaction~\cite{N3LO} at order
N$^3$LO, it neglects the contributions of any 3NFs.

This Letter is organized as follows. First, we briefly introduce
spherical coupled-cluster theory. Second, we compute the binding
energies of the nuclei $^4$He, $^{16}$O, $^{40}$Ca, $^{48}$Ca, and
$^{48}$Ni with the ``bare'' chiral NN potential. 

{\it Spherical coupled-cluster theory.} Coupled-cluster
theory~\cite{Coe58,Coe60,Ciz66,Ciz69,Kuem78,Bar07} is based on the
similarity transform
\be
\label{hbar}
\overline{H}=e^{-\hat{T}} \hat{H} e^{\hat{T}}
\ee
of the normal-ordered Hamiltonian $\hat{H}$. Here, the Hamiltonian is
normal-ordered with respect to a product state $|\phi\rangle$ which
serves as a reference. Likewise, the
particle-hole cluster operator
\begin{equation}
\label{T}
\hat{T} = \hat{T}_1 + \hat{T}_2 + \ldots + \hat{T}_A
\end{equation}
is defined with respect to the reference state. The $k$-particle 
$k$-hole ($k$p-$k$h) cluster operator is
\begin{equation}
\hat{T}_k =
\frac{1}{(k!)^2} \sum_{i_1,\ldots,i_k; a_1,\ldots,a_k} t_{i_1\ldots
i_k}^{a_1\ldots a_k}
\hat{a}^\dagger_{a_1}\ldots\hat{a}^\dagger_{a_k}
\hat{a}_{i_k}\ldots\hat{a}_{i_1} \,.
\end{equation}
Here and in the following, $i, j, k,\ldots$ label occupied
single-particle orbitals, while $a,b,c,\ldots$ label unoccupied
orbitals of the reference state, i.e. it should have significant
overlap with the ground state. Throughout this work we will restrict
ourselves to the CCSD approximation $\hat{T}\approx
\hat{T}_1+\hat{T}_2$. The unknown amplitudes $t_i^a$ and $t_{ij}^{ab}$
in Eq.~(\ref{T}) are determined from the solution of the
coupled-cluster equations
\begin{eqnarray}
\label{ccsd1}
0 &=& \langle \phi_i^a | \overline{H} | \phi\rangle \, \\
\label{ccsd2}
0 &=& \langle \phi_{ij}^{ab} | \overline{H} | \phi\rangle \,.
\end{eqnarray}
Here $|\phi_i^a\rangle = \hat{a}_{a}^\dagger\hat{a}_{i}|\phi\rangle$
is a 1p-1h excitation of the reference state, and
$|\phi_{ij}^{ab}\rangle$ is a similarly defined 2p-2h excited
state. The CCSD equations~(\ref{ccsd1}) and (\ref{ccsd2}) thus demand
that the reference state $|\phi\rangle$ is an eigenstate of the
similarity transformed Hamiltonian~(\ref{hbar}) in the space of all
1p-1h and 2p-2h excited states. Once the CCSD equations are solved,
the correlation energy of the ground state is computed as
\be 
E_{\rm corr} = \langle \phi | \overline{H} | \phi\rangle \,.  
\ee 

Coupled-cluster theory fulfills Goldstone's linked cluster theorem and
therefore yields size-extensive results. This is particularly
important in applications to medium-mass nuclei.  Within the CCSD
approximation, the computational effort scales as $n_o^2 n_u^4$ where
$n_o$ and $n_u$ denote the occupied and unoccupied orbitals of the
reference state $|\phi\rangle$, respectively. Thus, the computational
effort is much smaller than within the configuration interaction for a
given model space. This method has recently been employed in several
{\it ab initio} nuclear structure
calculations~\cite{Hei99,Mi00,Dean04,Kow04,Wlo05}. It is also able
to compute lifetimes of unstable nuclei~\cite{Hag06}, to treat
3NFs~\cite{Hag07}, and it meets benchmarks~\cite{CCbench}.

For spherical reference states (i.e. nuclei with closed major shells
or closed subshells), one can employ the spherical symmetry to further
reduce the number of unknowns (i.e. the number of cluster
amplitudes). For such nuclei, the cluster operator~(\ref{T}) is a
scalar under rotation, and depends only on reduced amplitudes.  A
naive estimate shows that a model space of $n_o+n_u$ single-particle
states consists of only $(n_o+n_u)^{2/3}$ $j$-shells. Thus, the entire
computational effort is approximately reduced by a power $2/3$ within
the spherical scheme compared to the $m$-scheme. We have derived and
implemented the spherical scheme within the CCSD approximation. We
tested that our $m$-scheme code and the spherical code give identical
results for several cases.

{\it Results.}  The single-particle basis consists of wave functions
of the spherical harmonic oscillator with the spacing $\hbar\omega$,
the radial quantum number $n$, and angular momentum $l$, and we include
single-particle states with $2n+l\le N$ in our model space. The
largest model space we consider ($N=14$) consists of 15 oscillator
shells. In such a large model space, configuration interaction becomes
impossible as the proton space alone consists of about $10^{40}$
Slater determinants for $^{40}$Ca. We first transform the Hamiltonian
to the spherical Hartree-Fock basis, and the CCSD equations are solved
in this basis.  Fully converged observables must be independent of the
parameters $N$ and $\hbar\omega$ of our single-particle basis. In practice,
we cannot go to infinitely large spaces, and the dependence of our
results on these parameters serve to gauge the convergence.

\begin{figure}[h]
\includegraphics[width=0.35\textwidth,clip=]{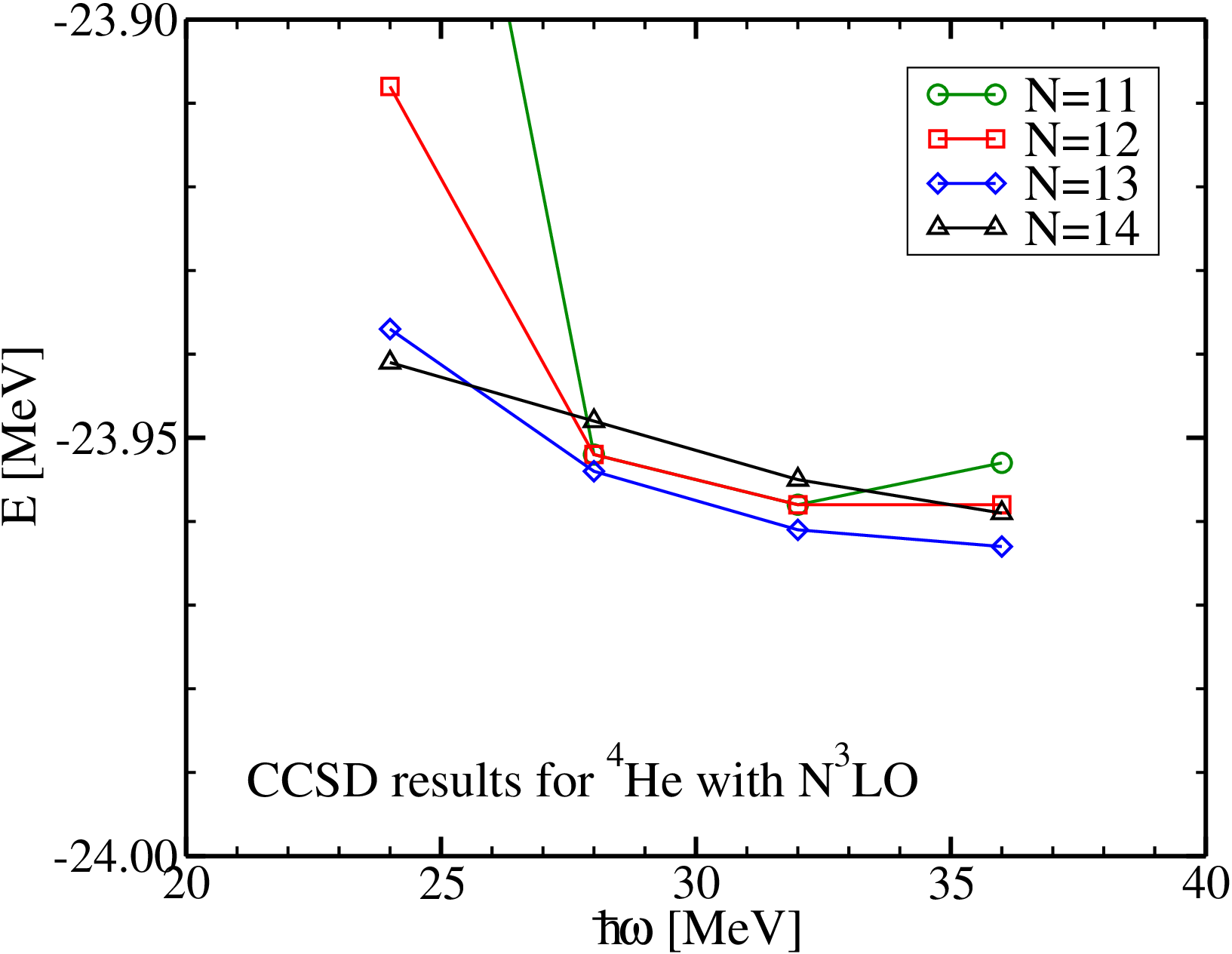}
\caption{(Color online) CCSD binding energy for $^4$He from the chiral
 NN potential at order N$^3$LO as a function of the oscillator spacing
 $\hbar\omega$ and the size of the model space.}
\label{fig1}
\end{figure}

As a test case, Fig.~\ref{fig1} shows that the CCSD results for $^4$He
are converged within a few keV with respect to increases in the size
of the model space (denoted by $N$) and variation of the oscillator
frequency. The triples corrections are not yet available within the
spherical scheme, and we employ our $m$-scheme code for this
purpose. The CCSD-T1 triples correction~\cite{ccsdt-n} in
model spaces up to $N=7$ yields another 1.3~MeV of binding. Thus, the
CCSD-T1 results are very close to the virtually exact
Faddeev-Yakubowski result $E=-25.41$~MeV quoted in
Ref.~\cite{Caurier04} for the same chiral NN interaction. The
experimental value is $E=-28.3$~MeV, and the additional binding is due
to the missing 3NFs.

\begin{figure}[h]
\includegraphics[width=0.35\textwidth,clip=]{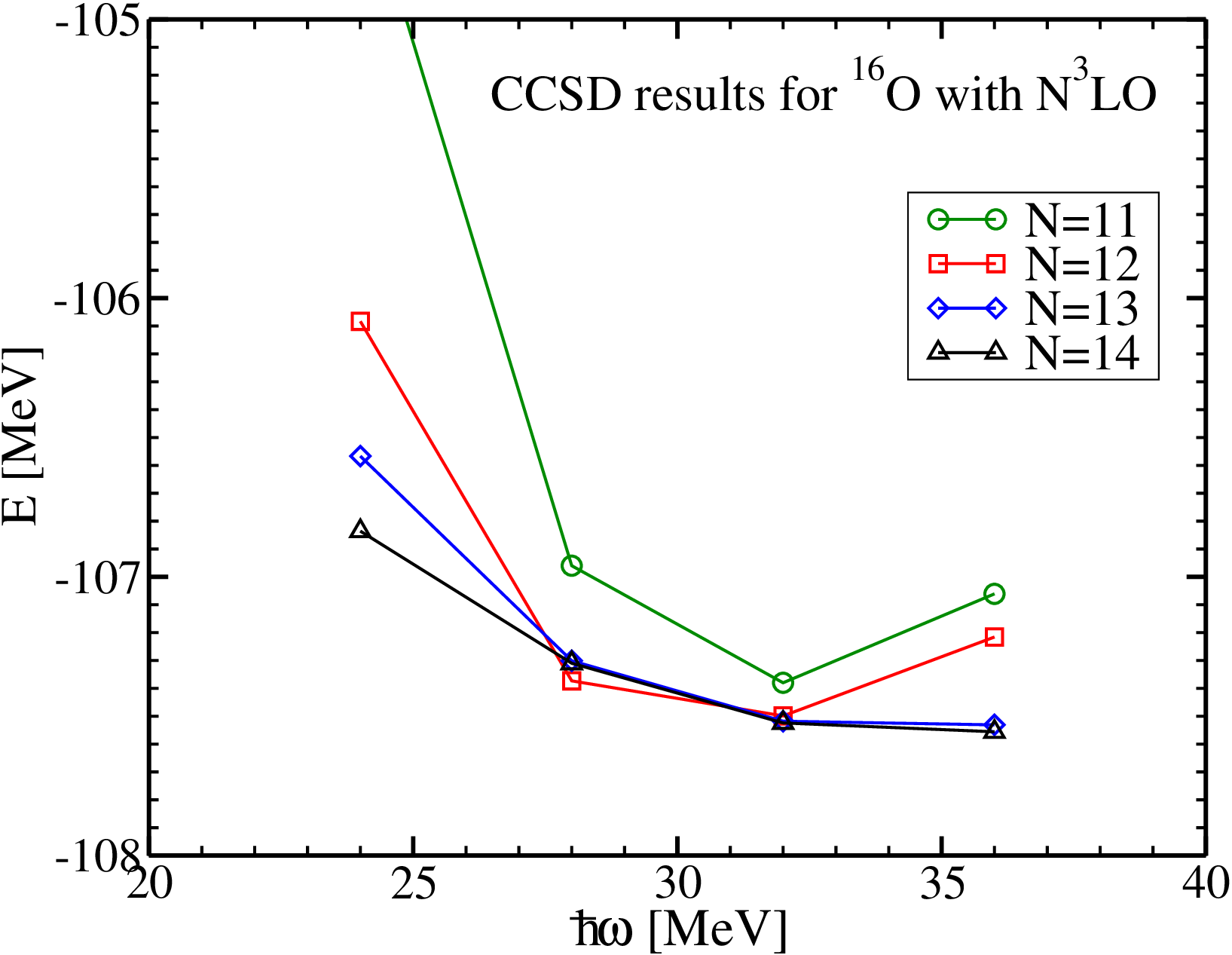}
\caption{(Color online) CCSD binding energy for $^{16}$O from the
 chiral NN potential at order N$^3$LO as a function of the oscillator
 spacing $\hbar\omega$ and the size of the model space.}
\label{fig2}
\end{figure}

The CCSD energies for $^{16}$O (see Fig.~\ref{fig2}) are converged
within the order of about 100~keV and change by less than 1~MeV over a
considerable variation of the oscillator frequency. This result is in
reasonably good agreement with the work by Fujii {\it et al.} who obtained
-110~MeV as the binding energy from the UOMA~\cite{Fujii04}. Recall that both methods are approximations and based on similarity-transformed Hamiltonians. 

\begin{figure}[h]
\includegraphics[width=0.35\textwidth,clip=]{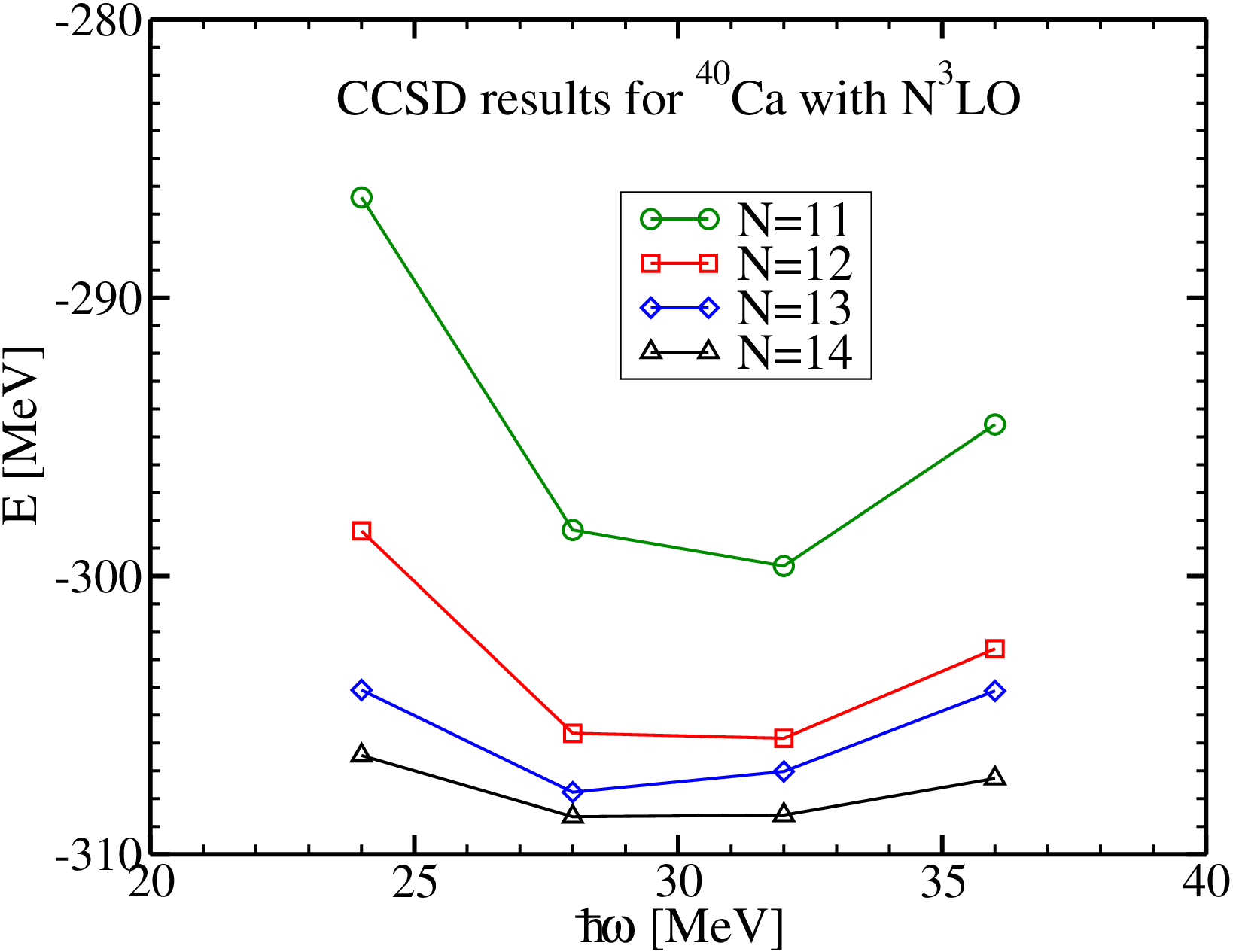}
\caption{(Color online) CCSD binding energy for $^{40}$Ca from the
 chiral NN potential at order N$^3$LO as a function of the oscillator
 spacing $\hbar\omega$ and the size of the model space.}
\label{fig3}
\end{figure}

We turn to nuclei in the mass-40 region.  The CCSD results for
$^{40}$Ca are shown in Fig.~\ref{fig3}. Increasing the model space
from $N=13$ to $N=14$ yields an additional 0.9~MeV, and the
$\hbar\omega$-dependence is less than 2.2~MeV over the considered
range of oscillator frequencies. Thus, the convergence with respect to
the parameters of our model space is very satisfactory, and we are missing 
about 10\% of the experimental binding energy of -342MeV.

\begin{figure}[h]
\includegraphics[width=0.35\textwidth,clip=]{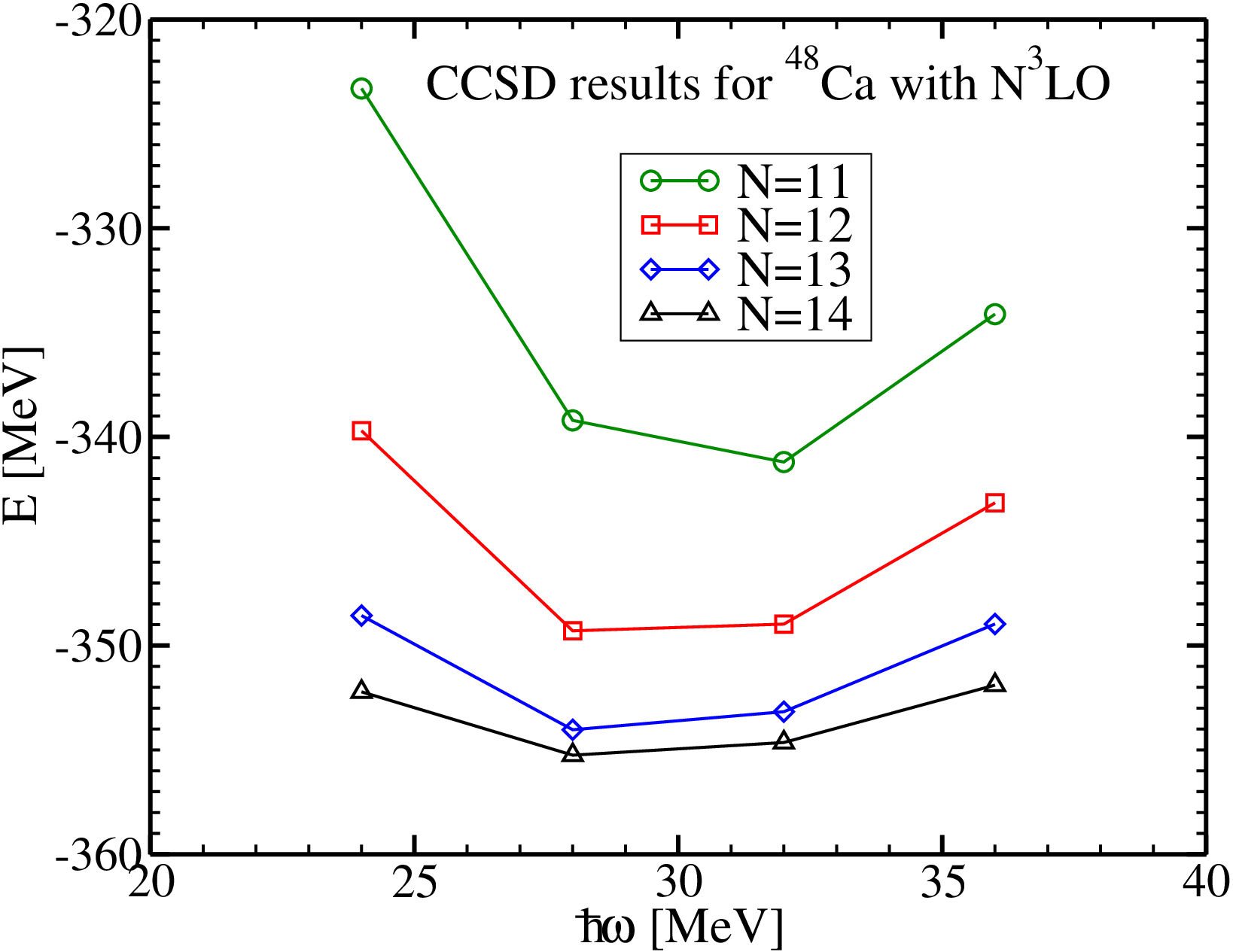}
\caption{(Color online) CCSD binding energy for $^{48}$Ca from the
 chiral NN potential at order N$^3$LO as a function of the oscillator
 spacing $\hbar\omega$ and the size of the model space.}
\label{fig4}
\end{figure}

\begin{figure}[h]
\includegraphics[width=0.35\textwidth,clip=]{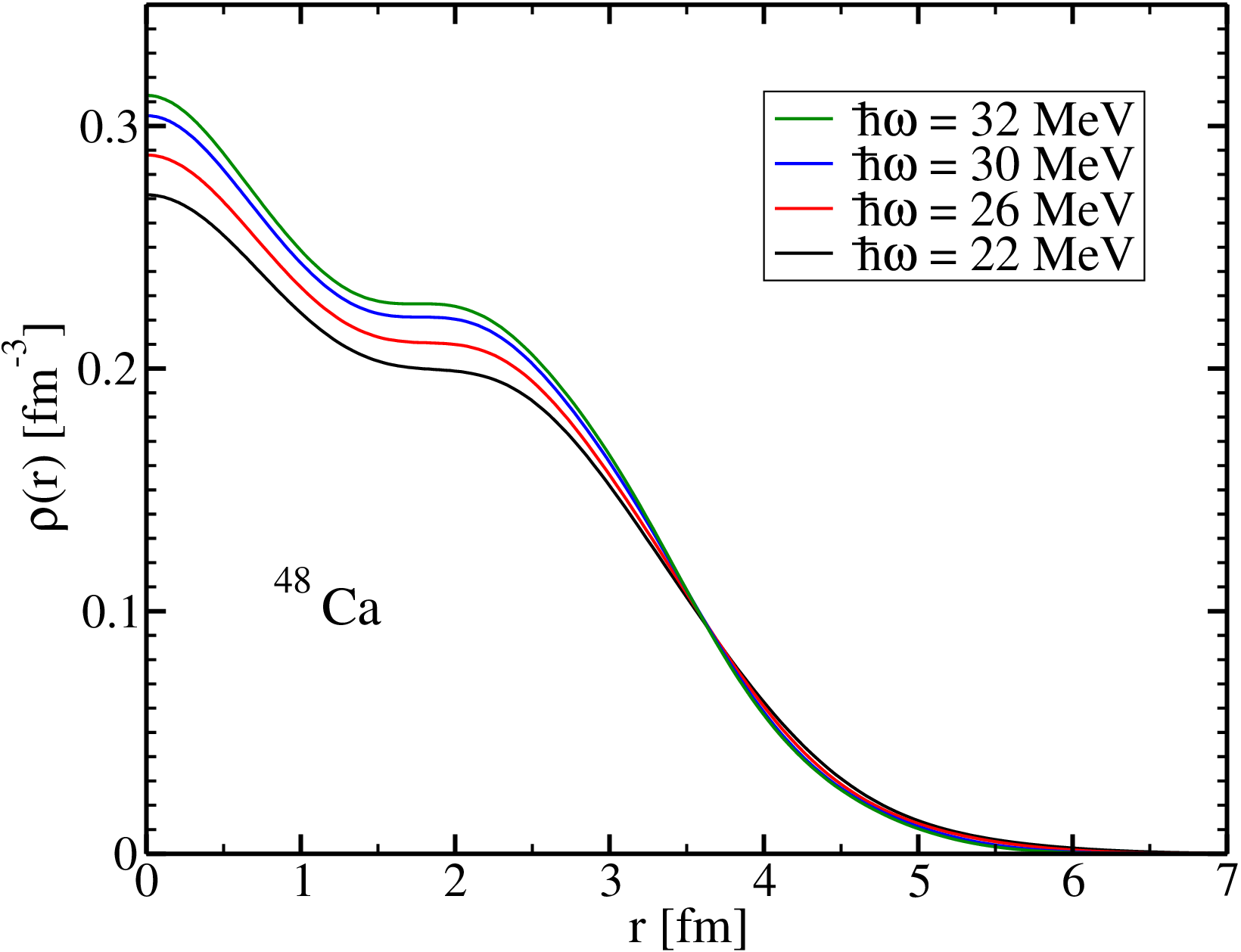}
\caption{(Color online) Densities for $^{48}$Ca from the chiral NN
potential at order N$^3$LO for four different values of the oscillator
spacing $\hbar\omega$.}
\label{fig5}
\end{figure}

We also computed the binding energy of the mirror nuclei $^{48}$Ca and
$^{48}$Ni. For $^{48}$Ca the convergence of the results is
satisfactory as shown in Fig.~\ref{fig4}, and the convergence is very
similar for $^{48}$Ni.  $^{48}$Ni was discovered only
recently~\cite{Ni48}. It is believed to be a two-proton emitter, and
its lifetime is very large compared to a typical nuclear time scale
(i.e. the ``orbital period'' of a nucleon inside the nucleus). Thus,
we can describe $^{48}$Ni in terms of a spherical Hartree-Fock basis
based on the oscillator orbitals. Recall that the chiral interaction
includes charge symmetry-breaking and charge independence-breaking
effects, and we also included the Coulomb interaction. The difference
of our CCSD results for the mirror nuclei $^{48}$Ca and $^{48}$Ni is
1.38~MeV per nucleon and stems from these combined effects.
Theoretical mass table evaluations~\cite{Audi} suggest that the
binding energy of $^{48}$Ni is 1.43~MeV per nucleon smaller than for
$^{48}$Ca. Our results are in good agreement with this estimate. The
density of $^{48}$Ca is shown in Fig.~\ref{fig5}. The results still
exhibit a dependence of the oscillator spacing $\hbar\omega$, and the
central density decreases with decreasing $\hbar\omega$. This
observable is less well converged than the energy with respect to the
size of the model space. The convergence is slow with respect to the
maximum radial quantum number $n$ employed in our model space, while the
single-particle angular momentum $l$ could be limited to $l\le 7$.

Table~\ref{tab1} summarizes some of our results which are taken at
$\hbar\omega=28$MeV in the largest model spaces. We computed the
potential energy $V$ via the Hellman-Feynman theorem. The fourth
column shows the energy deviation $\Delta E\equiv E-E_{\rm exp}$ from
the experimental binding energy $E_{\rm exp}$ for the considered
nuclei. This difference is mainly due to the omitted 3NFs and the
missing triples correction.  Note that $^{40}$Ca is particularly
tightly bound when compared to the other nuclei. The isotopes
$^{16}$O, $^{48}$Ca, and $^{48}$Ni all lack about $\Delta E/A\approx
1.2$~MeV of binding energy when compared to experiment, while this
difference is considerably smaller for $^{40}$Ca. This result is
somewhat surprising since $^{48}$Ca is thought to be a better example
of a doubly magic nucleus than $^{40}$Ca. There seems to be a
cancellation between triples corrections and contributions of 3NFs in
$^{40}$Ca. In other words, the isospin-dependence and/or
mass-dependence of the 3NF is expected to be non-trivial. The charge
radii are corrected according to Ref.~\cite{Friar} to account for the
finite charge radii of the nucleons. They are computed from the
leading approximation of the center-of-mass corrected intrinsic
density~\cite{Mi00}. Note that the radii change about 0.1-0.25 fm as the
oscillator spacing $\hbar\omega$ is varied in the range that is shown
in the previous figures, and they decrease with increasing values of
$\hbar\omega$.

\begin{table}[h]
\begin{tabular}{|c||r|r|r|r|r|}\hline
  Nucleus   & $E/A$   & $V/A$  & $\Delta E / A$ & $R$ & $R_{\rm exp}$\\\hline\hline
  $^{4}$He  &  -5.99  & -22.75 & 1.08 & 1.86 & 1.64\\\hline
  $^{16}$O  &  -6.72  & -30.69 & 1.25 & 2.71 & 2.74\\\hline
  $^{40}$Ca &  -7.72  & -36.40 & 0.84 & 3.24 & 3.48\\\hline
  $^{48}$Ca &  -7.40  & -37.97 & 1.27 & 3.22 & 3.47\\\hline
  $^{48}$Ni &  -6.02  & -36.04 & 1.21 & 3.50 &     \\\hline
\end{tabular}
\caption{CCSD results for various nuclei from the chiral N$^3$LO
nucleon-nucleon potential. The binding energy per nucleon, and
potential energy per nucleon are denoted as $E/A$ and $V/A$,
respectively. $\Delta E$ denotes the difference to the experimental
binding energy (difference to theoretical mass table evaluations for
$^{48}$Ni). $R$ and $R_{\rm exp}$ denote the computed and measured
charge radius. Energies are in units of MeV, and lengths in units of
fm.}
\label{tab1}  
\end{table}

We also compute the one-body density matrices $\rho_{pq}=\langle
\hat{a}^\dagger_p\hat{a}_q \rangle$ of the ground states
within the equation-of-motion CCSD~\cite{eomccsd}. The diagonalization of
this matrix yields natural orbitals and the corresponding
occupations. These model-dependent quantities are, of course, not
observables but rather tied to the specific interaction we
employed. The dominant occupation probabilities are larger than 0.95,
and this indicates that the considered nuclei are indeed doubly magic.
This result is non-trivial. Note that the Hartree-Fock approximation
does not even yield bound nuclei. Yet the CCSD correlations imprinted
onto the Hartree-Fock state yield a rather simple state. To our
knowledge, this is the first time the phenomenological shell-model
picture of independent nucleon motion arises within an {\it ab-initio}
approach.

{\it Summary.} We have studied the saturation properties of chiral NN
interactions at the order N$^3$LO in medium-mass nuclei within the
CCSD approximation of coupled-cluster theory. Our results exhibit a
very satisfactory convergence with respect to the size of the model
space and are only weakly dependent on the oscillator parameter. We
find that the ``bare'' chiral NN potential underbinds nuclei by about
1~MeV per nucleon. The comparison of $^{40}$Ca with $^{48}$Ca and
$^{48}$Ni hints at an isospin dependence of the 3NF in medium-mass
nuclei. Within the CCSD approximation, the proton-rich nucleus
$^{48}$Ni is less tightly bound by 1.38~MeV per nucleon than its
mirror nucleus $^{48}$Ca, and this result is in good agreement with
theoretical mass table evaluations. These calculations pave the
way to probing chiral interactions in even heavier nuclei and 
link the phenomenological shell model to ab-initio calculations.

We thank S.K.~Bogner, R.~J.~Furnstahl, and A.~Schwenk for useful
discussions.  This work was supported by the U.S. Department of Energy
under Contract Nos. \ DE-AC05-00OR22725 with UT-Battelle, LLC (Oak
Ridge National Laboratory), and DE-FC02-07ER41457 (University of
Washington), and under Grant No.\ DE-FG02-96ER40963 (University of
Tennessee). This research used resources of the National Center for
Computational Sciences at Oak Ridge National Laboratory.

\end{document}